\documentclass{elsarticle}

\usepackage{graphicx}
\usepackage{balance}  
\usepackage{amsfonts}
\usepackage{array}
\usepackage{footnote}
\usepackage{url}
\usepackage{multirow}
\usepackage{subfig}
\usepackage{xspace}

\newcommand{\naivepart}{\textsc{NaivePart}\xspace}
\newcommand{\incrementalupdate}{\textsc{IncrementalPart}\xspace}
\newcommand{\bladyg}{\textsc{bladyg}\xspace}
\newcommand{\Degree}{d}
\newcommand{\Block}{\textit{block}}
\newcommand{\Core}{k}
\newcommand{\adj}{\textit{adj}}

\newcommand{\workerCompute}{\textsf{workerCompute}\xspace}
\newcommand{\masterCompute}{\textsf{masterCompute}\xspace}

\newcommand{\akka}{\textsc{akka}\xspace}
\newcommand{\flume}{\textsc{Flume}\xspace}

\newcommand{\fluentd}{\textsc{Fluentd}\xspace}
\newcommand{\ubupdate}{\textsc{UB-Update}\xspace}
\newcommand{\dynamicdfep}{\textsc{DynamicDFEP}\xspace}

\newtheorem{definition}{Definition}
\newtheorem{theorem}{Theorem}

\begin{document}


\begin{frontmatter}

\title{BLADYG: A Graph Processing Framework \\for Large Dynamic Graphs} 

\author[loria]{Sabeur Aridhi\corref{cor1}}
\ead{sabeur.aridhi@loria.fr}

\author[trento]{Alberto Montresor}
\ead{alberto.montresor@unitn.it}

\author[trento]{Yannis Velegrakis}
\ead{velgias@disi.unitn.eu}

\cortext[cor1]{Corresponding author}

\address[loria]{University of Lorraine, LORIA, Campus Scientifique, BP 239, 54506 Vandoeuvre-l\`es-Nancy, France}

\address[trento]{University of Trento, Italy}

%
%
%
\sloppy
\balance
\begin{abstract}
Recently, distributed processing of large dynamic graphs has become very popular, especially in certain domains such as social network analysis, Web graph analysis and spatial network analysis. 
In this context, many distributed/parallel graph processing systems have been proposed, such as Pregel, GraphLab, and Trinity. 
These systems can be divided into two categories: (1) vertex-centric and (2) block-centric approaches. 
In vertex-centric approaches, each vertex corresponds to a process, and message are exchanged among vertices. 
In block-centric approaches, the unit of computation is a block, a connected subgraph of the graph, and message exchanges occur among blocks.  
In this paper, we are considering the issues of scale and dynamism in the case of block-centric approaches. 
We present \bladyg, a block-centric framework that addresses the issue of dynamism in large-scale graphs. 
We present an implementation of \bladyg on top of \akka framework. We experimentally evaluate the performance of the proposed framework.
\end{abstract}



\begin{keyword}
Distributed graph processing, Dynamic graphs, \akka framework
\end{keyword}


\end{frontmatter}
\section{Introduction}
In the last decade, the field of distributed processing of large-scale graphs has attracted considerable attention~\cite{bdr}. 
This attention has been motivated not only by the increasing size of graph data, but also by its huge number of applications, such as the analysis of social networks~\cite{5992567}, web graphs~\cite{Alvarez} and spatial networks~\cite{JTLU23}. 
In this context, many distributed/parallel graph processing systems have been proposed, such as Pregel~\cite{pregel}, GraphLab~\cite{graphlab}, and Trinity~\cite{trinity}. These systems can be divided into two categories: (1) vertex-centric and (2) block-centric approaches. 
Vertex-centric approaches divide input graphs into partitions, and employ a "think like a vertex" programming model to support iterative graph computation~\cite{pregel,Tian:2013}. Each vertex corresponds to a process, and message are exchanged among vertices. 
In block-centric approaches~\cite{blogel}, the unit of computation is a block -- a connected subgraph of the graph -- and message exchanges occur among blocks. 
The vertex-centric approaches have been proved to be useful for many graph algorithms. However, they do not always perform efficiently, because they ignore the vital information about graph partitions, which represent a real subgraph of the original input graph, instead of a collection of unrelated vertices. 

In our work, we are considering the issues of scale and dynamism in the case of block-centric approaches~\cite{bladyg}. 
Particularly, we are considering big graphs known by their evolving and decentralized nature. 
For example, the structure of a big social network (e.g., Twitter, Facebook) changes over time (e.g., users start new relationships and communicate with different friends). 


We present  \bladyg, a block-centric framework that addresses the issue of dynamism in large-scale graphs.  
\bladyg can be used not only to compute common properties of large graphs, but also to maintain the computed properties when new edges and nodes are added or removed. 
The key idea is to avoid the re-computation of graph properties from scratch when the graph is updated. 
\bladyg limits the re-computation to a small subgraph depending on the undertaken task. 
We present a set of abstractions for \bladyg that can be used to design algorithms for any distributed graph task. 

More specifically, our contributions are: 
\begin{itemize}
\item We introduce \bladyg and its computational distributed model.
\item We present an implementation of \bladyg on top of \akka~\cite{akka}, a framework for building highly concurrent, distributed, and resilient message-driven applications.
\item We experimentally evaluate the performance of the proposed framework, by applying it to problems such as distributed $k$-core decomposition of large graphs and partitioning of large dynamic graphs. 
\end{itemize}


The rest of the paper is organized as follows. 
In Section \ref{related}, we highlight existing works on distributed graph processing on large and dynamic graphs.  
In Section \ref{algo}, we present the system overview of \bladyg. 
In Section \ref{app}, we present some research problems that can be solved using \bladyg. 
Finally, we describe our experimental evaluation in Section \ref{exp}.

\section{Related works}
\label{related}
In this section we highlight the relevant literature in the field of large graph processing. 
We consider two kinds of frameworks: (1) graph processing frameworks and (2) frameworks for the processing of large and dynamic graphs. 

\paragraph*{Graph processing frameworks} 

\begin{figure}[h]
\centering

 \includegraphics[width=0.9\textwidth]{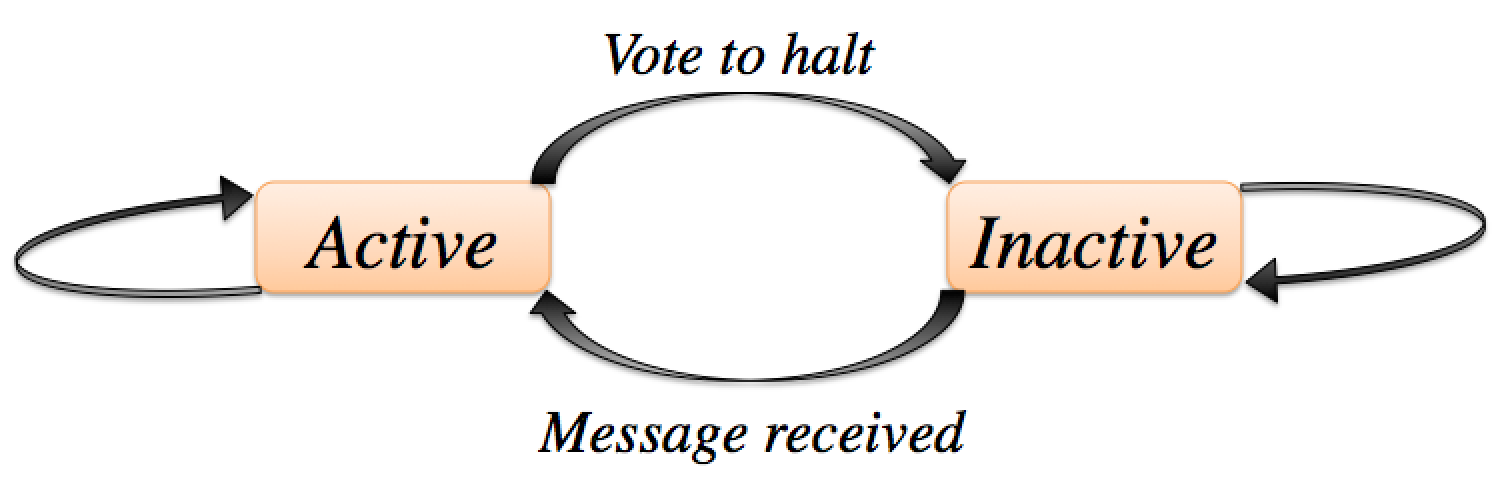}

\caption{Vertex's state machine in Pregel.}
\label{pregel}
\end{figure}

Pregel \cite{pregel} is a computational model for large-scale graph processing problems. 
In Pregel, message exchanges occur among vertices of the input graph. As shown in Figure \ref{pregel}), each vertex is associated to a state that controls its activity.  

Each vertex can decide to halt its computation, but can be woken up at every point of the execution by an incoming message. 
At each superstep of the computation a user defined vertex program is executed for each active vertex. 
The user defined function will take the vertex and its incoming messages as input, change the vertex value and eventually send messages to other vertices through the outgoing edges. 

\begin{figure}[t]

\centering
 \includegraphics[width=0.9\textwidth]{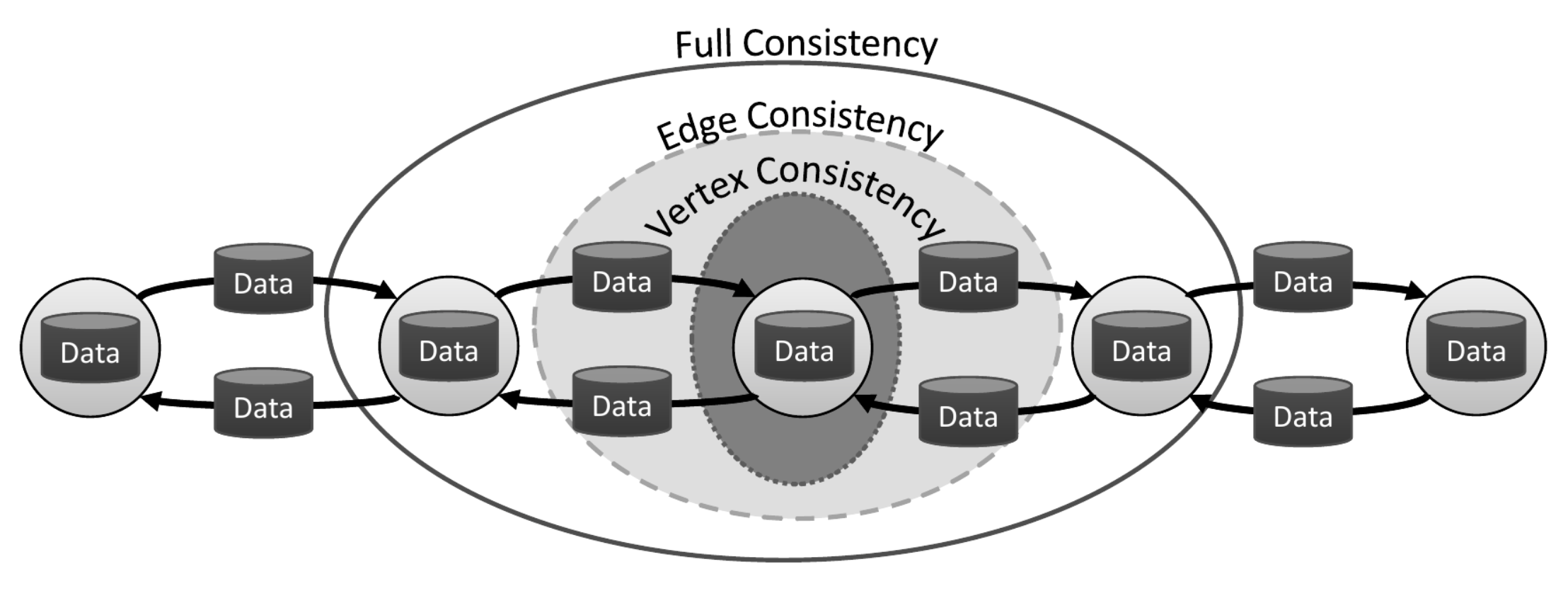}

\caption{View of the scope of a vertex in GraphLab.}
\label{scope}
\end{figure}

GraphLab \cite{graphlab} is a graph processing framework that share the same motivation with Pregel. 
While pregel targets Google'€™s large distributed system, GraphLab addresses shared memory parallel systems which means that there is more focus on parallel access of memory than on the issue of efficient message passing and synchronization. 
In the programming model of GraphLab, the users define an update function that can change all data associated to the scope of that node (its edges or its neighbors). 
Figure \ref{scope} shows the scope of a vertex: an update function called on that vertex will be able to read and write all data in its scope. 
We notice that that scopes can overlap, so simultaneously executing two update functions can result in a collision. 
In this context, GraphLab offers some consistency models in order to allow its users to trade off performance and consistency as appropriate for their computation. As described in Figure \ref{scope}, GraphLab offers a fully consistent model, a vertex consistent model or an edge consistent model. 


Powergraph \cite{Gonzalez:2012:PDG:2387880.2387883} is an abstraction that exploits the structure of vertex-programs and explicitly factors computation over edges instead of vertices
It uses a greedy approach, processing and assigning each edge before moving to the next.  
It keeps in memory the current sizes of each partition and, for each vertex, the set of partitions that contain at least one edge of that vertex. 
If both endpoints of the current edge are already inside one common partition, the edge will be added to that partition. 
If they have no partition in common, the node with the most edges still to assign will choose one of its partitions. 
If only one node is already in a partition, the edge will be assigned to that partition. Otherwise, if both nodes are free, the edge will be assigned to the smallest partition. 


GraphX \cite{Xin:2013:GRD:2484425.2484427} is a library provided by Spark \cite{Zaharia:2012:RDD:2228298.2228301}, a framework for distributed and parallel programming. 
Spark introduces Resilient Distributed Datasets (RDD), that can be split in partitions and kept in memory by the machines of the cluster that is running the system. 
These RDD can be then passed to one of the predefined meta-functions such as map, reduce, filter or join, that will process them and return a new RDD. 
In GraphX, graphs are defined as a pair of two specialized RDD. The first one contains data related to vertices and the second one contains data related to edges of the graph. 
New operations are then defined on these RDD, to allow to map vertices's values via user defined functions, join them with the edge table or external RDDs, or also run iterative computation. 

%
%

\paragraph*{Processing of large and dynamic graphs}
Chronos \cite{chronos} is an execution and storage engine designed for running in-memory iterative graph computation on evolving graphs. 
Locality is an important aspect of Chronos, where the in-memory layout of temporal graphs and the scheduling of the iterative computation on temporal and evolving graphs are carefully designed. The design of Chronos further explores the interesting interplay among locality, parallelism, and incremental computation in supporting common mining tasks on temporal graphs. We notice that traditional graph processing frameworks arrange computation around each vertex/edge in a graph; while temporal graph engines, in addition, calculate the result across multiple snapshots. 
Chronos makes a decision to batch operations associated with each vertex (or each edge) across multiple snapshots, instead of batching operations for vertices/edges within a certain snapshot \cite{chronos}. 

The problem of distributed processing of large dynamic graphs has attracted considerable attention. In this context, several traditional graph operations such as $k$-core decomposition and maximal clique computation have been extended to dynamic graphs \cite{DBLP:conf/bigdata/XuCFB14} \cite{6702486} \cite{Sariyuce:2013:SAK:2536336.2536344} \cite{DBLP:journals/tkde/LiYM14}. While this field of large dynamic graph analysis represent an emerging class of applications, it is not sufficiently addressed by the current graph processing frameworks and only specific graph operations have been studied in the context of dynamic graphs.

\section{The BLADYG framework}
In this section, we first describe \bladyg and its main components. Then, we give a running example that helps to understand the basic \bladyg operations. 

\subsection{BLADYG system overview}

\label{algo}

\begin{figure*}[t]
\centering
\includegraphics[width=0.9\textwidth]{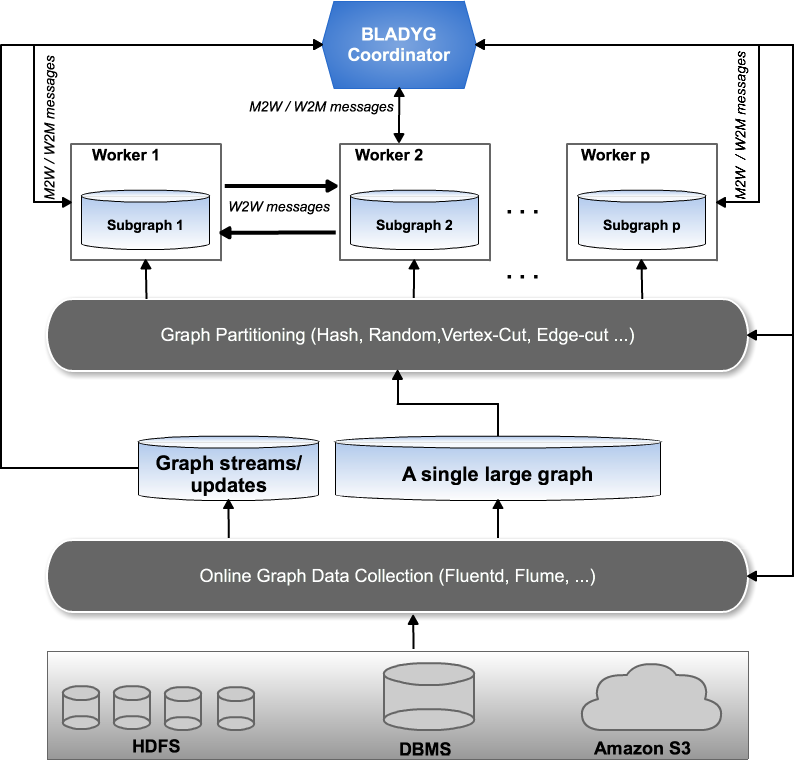}
\caption{\bladyg system overview}
\label{fig1}
\end{figure*}

Figure~\ref{fig1} provides an architectural overview of the \bladyg framework.
\bladyg starts its computation by collecting the graph data from various data sources including local files, Hadoop Distributed File System (HDFS) and Amazon Simple Storage Service (Amazon S3). In \bladyg, graph data collection can be done using existing open source collection tools including \flume \cite{flume} and \fluentd \cite{fluentd}.
After collecting the graph data, \bladyg partitions the input graph into multiple partitions, each of them assigned to a different worker. 
Each partition/block is a connected subgraph of the input graph. 
This partitioning step is performed by a partitioner worker that supports several types of predefined partitioning techniques such as hash partitioning, random partitioning, edge-cut and vertex-cut. 
\begin{itemize}
\item In hash partitioning, edges are distributed across machines according to a user-defined hash function. 
\item In random partitioning, edges are distributed across machines randomly. 
\item In vertex-cut, edges are evenly distributed across machines with the goal of minimizing the number of replicated vertices.
\item In edge-cut partitioning, the vertices of a graph are divided into disjoint clusters of nearly equal size, while the number of edges that span separated clusters is minimum. 
\end{itemize}
In addition to the provided partitioning techniques, \bladyg users may deploy existing graph partitioning techniques including Metis \cite{metis} and JaBeJa \cite{jabeja}. \bladyg users may also implement their own partitioning methods. 
It is important to mention that \bladyg allows to process large graphs that already distributed among a set of machines. 
This is motivated by the fact that the majority of the existing large graphs are already stored in a distributed way, either because they cannot be stored on a single machine due to their sheer size, or because they get processed and analyzed with  decentralized techniques that require them to be distributed among a collection of machines.
Each worker loads its block and performs both local and remote computations, after which the status of the blocks is updated. 
The coordinator/master worker orchestrates the execution of \bladyg in order to deal with incremental changes on the input data. 
Depending on the graph task, the coordinator builds an execution plan which consists of an ordered list of both local and distant computation to be executed by the workers. 

Each worker performs two types of operations: 
\begin{enumerate}
 \item \textbf{Intra-block computation}: in this case, the worker do local computation on its associated block (partition) and modifies either the status of the block and/or the states of the nodes inside the block. 
 \item \textbf{Inter-block computation}: in this case, the worker asks distant workers to do computation and after receiving the results it updates the status of its associated block. 
\end{enumerate}

\bladyg framework for large dynamic graph analysis operates in three computing modes: 
 In \textit{M2W-mode/W2M-mode}, message exchanges between the master and all workers are allowed.  
 The master uses this mode to ask a distant worker to look for candidate nodes i.e., nodes that need to be updated depending on the undertaken task. 
 The worker uses this mode to send the set of computed candidate nodes to the master. 
 In \textit{W2W-mode}, message exchanges between workers are allowed. 
 The workers use this mode in order to propagate the search for candidate nodes to one or more distant workers. 
 In \textit{Local-mode}, only local computation is allowed. 
 This mode is used by the worker/master to do local computation.  

A typical \bladyg computation consists of: (1) an input graph, (2) a set of incremental changes, (3) a sequence of worker/master operations and (4) an output. 
\begin{enumerate}
\item The input of \bladyg framework is an undirected graph. 
This graph is represented by a set of vertices and a set of edges. A vertex is defined by its unique ID and a value, whereas an edge is defined by its source ID, target ID, and value. 
\item Incremental changes or graph updates consists of edge/node insertions and/or removals. Graph updates are continuously read from the data sources using one of the data collection tools provided by \bladyg.  
\item A worker operation is a user-defined function that is executed by one or many workers in parallel depending on the logic of the graph task.  
Within each worker operation, the state of the associated block is updated and all the computing modes of \bladyg are activated. 
Within each master operation, a user defined function that defines the orchestration mechanism of the master is executed. During a master operation \textit{Local-mode} and \textit{M2W-mode} are activated. 
\item The output of a \bladyg program consists of an updated list of vertices and an updated list of edges. 
\end{enumerate}

\subsection{Illustrative example}
Here, we provide an illustrative example to explain the principle of our approach. 
 \begin{figure}[!h]
 \centering
 \includegraphics[width=.8\textwidth]{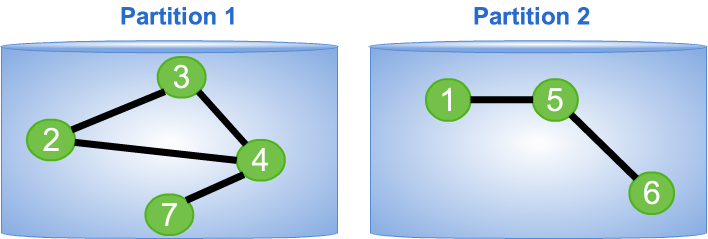}
 \caption{A graph example distributed into two partitions.}
 \label{fig:exp1}
 \end{figure}

\begin{figure}[!h]
 \centering
 \includegraphics[width=.8\textwidth]{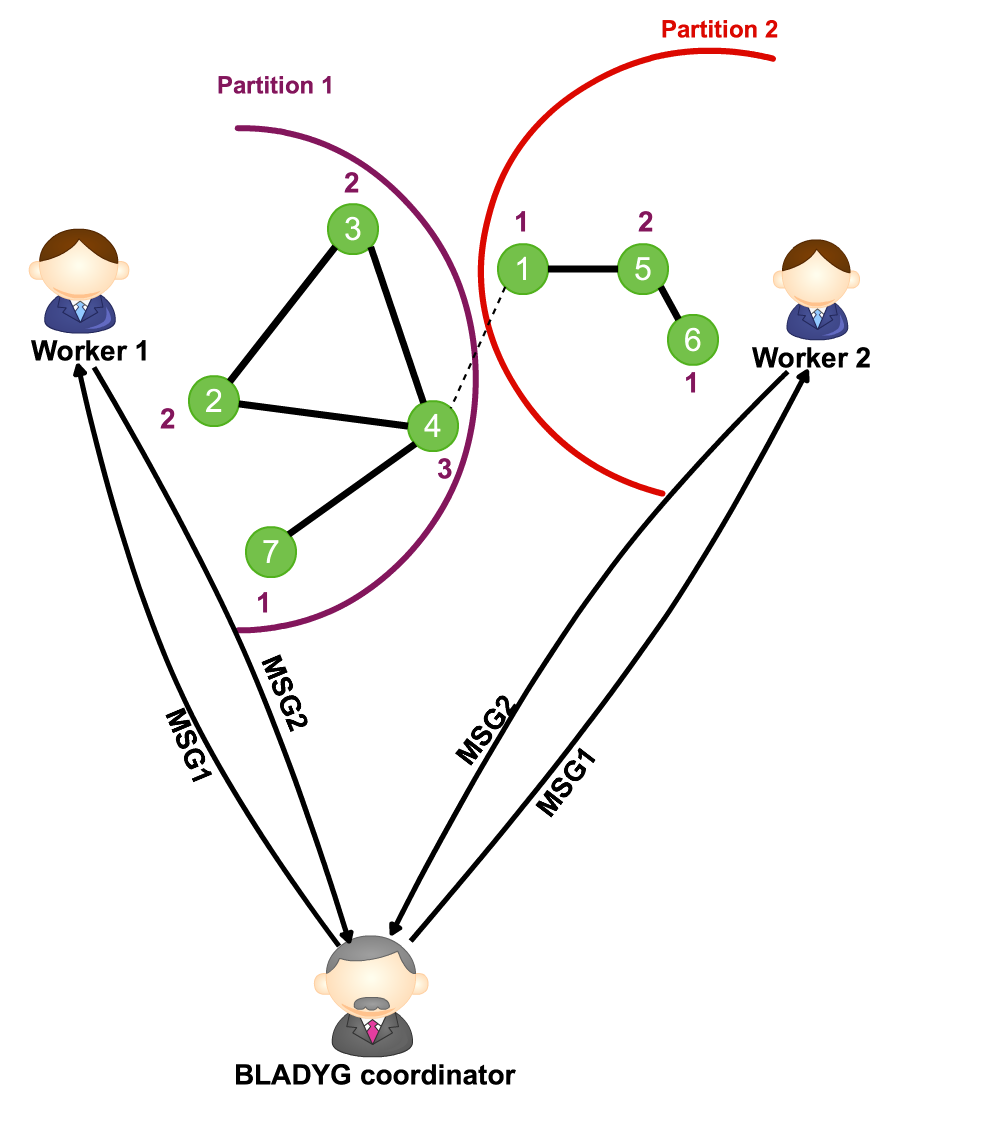}
 \caption{The sequence of messages exchanged among the coordinator and the worker nodes.}
 \label{fig:exp2}
 \end{figure}

Consider the graph $G=(V,E)$ included in Figure~\ref{fig:exp1}, and suppose that it is splitted in two partitions, each processed by a separate worker. We consider the task of computing the degree of all the nodes in $G$. The system is completed by the master node, as shown in Figure~\ref{fig:exp2}.

\begin{figure}[!h]
 \centering
 \includegraphics[width=.75\textwidth]{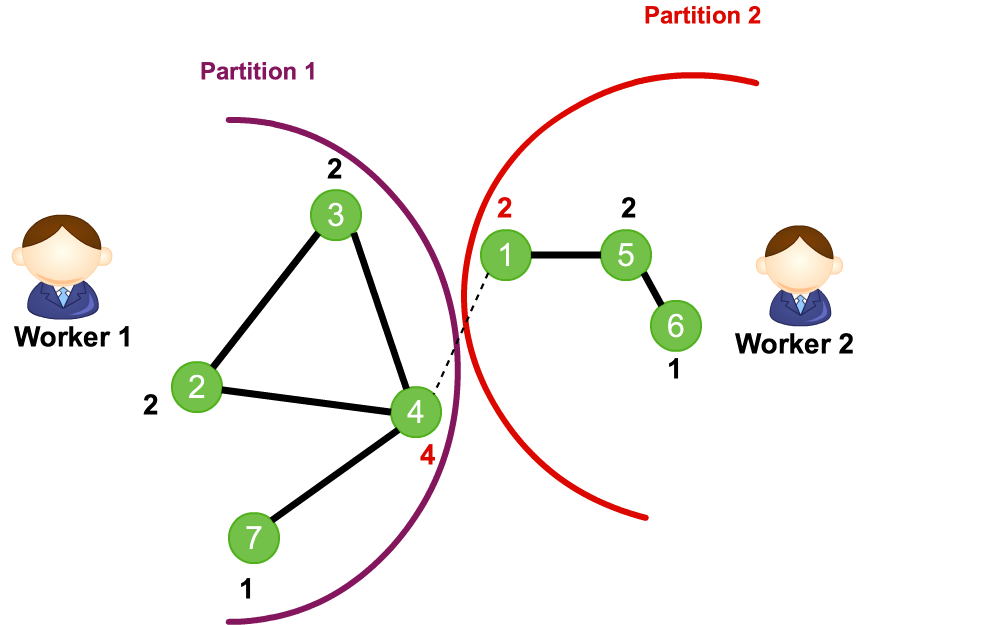}
 \caption{The updated graph.}
 \label{fig:exp3}
 \end{figure}

A \bladyg solution for computing the degree of all the nodes in a given graph consists of two steps. 
 The first step consists in executing several worker operations in order to compute the degree of nodes in all subgraphs in parallel.
 As a result of this step, the degree values of all the nodes of $G$ are computed. The degree values of the nodes of our graph example $G$ are presented in Figure~\ref{fig:exp2}. 
We assume that the incremental changes in our example consists of only one new edge that links node $4$ and node $1$. 
The second step of our \bladyg solution consists in selecting the set of nodes that need to be updated after considering the graph updates (insertion of the edge $(4,1)$).
In this example, only the nodes of the new edge need to be updated (nodes $1$ and $4$). The master sends a M2W message (MSG1) to worker $1$ (respectively to worker $2$) and asks the worker to increment the degree of node $4$ (respectively node $1$). The updated degree values of the nodes of our graph example $G$ are presented in Figure~\ref{fig:exp3}. 

After updating the degree of the node $4$ (respectively node  $1$), worker $1$ (respectively worker $2$) sends a notification message (MSG2) to the master. The master checks that all the graph updates were processed and stops the execution of the \bladyg program.

In this example, we only considered an insertion of a new edge between two existing nodes. It is important to mention that in real world applications, graph updates consists of insertion/deletion of several nodes/edges. We also mention that the complexity of the task of selecting the nodes that need to be updated after considering graph updates depends on the considered graph operation.

\section{Applications}
\label{app}
In this section, we apply \bladyg to solve some classic graph operations such as $k$--core decomposition~\cite{MontresorPM13}~\cite{debs2016a}, 
clique computation~\cite{DBLP:conf/bigdata/XuCFB14} 
and graph partitioning~\cite{DBLP:conf/europar/GuerrieriM15}~\cite{ideas2016a}. 
%

\subsection{Distributed $k$-core decomposition}
Let $G=(V,E)$ be an undirected graph  with $n=|V|$ nodes and $m=|E|$
edges. $G$ is partitioned into $p$ disjoint partitions $\{ V_1, \ldots, V_p \}$; in other words, $V = \cup_{i=1}^p V_i$ and $V_i \cap V_j = \emptyset$ for each $i,j$ such that $1 \leq i,j \leq p$ and $i \neq j$. 
The task of \emph{$k$--core decomposition}~\cite{batagely03} is condensed in the following two definitions: 

\begin{definition}
A subgraph $G(C)$ induced by the set $C \subseteq V$ is a \emph{$k$-core} if and
only if $\forall u \in C: \Degree_{G(C)}(u) \geq k$, and $G(C)$ is maximal, 
i.e., for each $\overline C \supset C$, there exists $v \in \overline C$ such 
that  $\Degree_{G(\overline C)}(v) < k$.
\end{definition}

\begin{definition}
A node in $G$ is said to have \emph{coreness} $k$ ($\Core_G(u)=k)$ if and only 
if it belongs to the $k$-core but not the $(k+1)$-core.
\end{definition}

%
A $k$-core of a graph $G=(V,E)$ can be obtained by recursively removing all the vertices of degree less than $k$, until all vertices in the remaining graph have degree at least $k$.  
The issue of distributed \emph{$k$--core decomposition} in dynamic graphs consists in updating the coreness of the nodes of $G$ when new nodes/edges are added and/or removed. 

\bladyg solves the problem of distributed \emph{$k$--core decomposition} in two steps. 
The first step consists in executing a $\workerCompute()$ operation that computes the coreness inside each of the blocks. 
Inside a block, each vertex is associated with $\Block(u)$, $\Degree_G(u)$ and $\Core_G(u)$, denoting the block of $u$, the degree and the coreness of $u$ in $G$, respectively. 
The second step consists in maintaining the coreness values after considering the incremental changes. 
Whenever a new edge $(u, v)$ is added to the graph, \bladyg first activates the \textit{M2W-mode} and computes the set of candidate nodes i.e., nodes whose coreness needs to be updated. 
This is done by two $\workerCompute()$ operations inside the workers that hold $u$ and $v$. 
The $\workerCompute()$ operations exploit Theorem~\ref{theorem1}, first stated and demonstrated by Li, Yu and Mao~\cite{DBLP:journals/tkde/LiYM14}, that identifies what are the \emph{candidate nodes} that may need to be updated whenever we add an edge: 

\begin{theorem} 
\label{theorem1}	
Let $G = (V,E)$ be a graph and $(u, v)$ be an edge to be inserted in $E$, with $u,v \in V$. A node $w \in V$ is said to be a \emph{candidate} to be updated based on the following three cases:
\begin{itemize} 
\item If $\Core(u) > \Core(v)$, $w$ is candidate if and only if $w$ is $k$-reachable from $v$ in the original graph $G$ and $k=\Core(u)$;
\item If $\Core(u) < \Core(v)$, $w$ is candidate if and only if $w$ is $k$-reachable from $u$ in the original graph $G$ and $k=\Core(v)$;
\item If $\Core(u) = \Core(v)$, $w$ is candidate if and only if $w$ is $k$-reachable from either $u$ and $v$ in the original graph $G$ and 
$k=\Core(u)$.
\end{itemize}
\end{theorem}

A node $w$ is $k$-\emph{reachable} from $u$ if $w$ is reachable from $u$ in the $k$-core of $G$; i.e., if there exists a path between $u$ and $w$ in the original graph such that all nodes in the path (including $u$ and $w$) have coreness equal to $k=\Core(u)$.

We notice that the executed $\workerCompute()$ operations may activate the \textit{W2W-mode} since the set of nodes to be updated may span multiple blocks/partitions. 
The nodes identified as potential candidates are sent back to the coordinator node that orchestrates the execution and computes, by executing a $\masterCompute()$ operation, the correct coreness values of the candidate nodes. 
%

 \subsection{Distributed edge partitioning}
 
 Edge partitioning is a classical problem in graph processing in which edges of a given graph, rather than its vertices, are partitioned into disjoint subsets. 
 Given a graph $G = (V, E)$ and a parameter $K$, an edge partitioning of $G$ subdivides all edges into a collection $E_1, \cdots, E_K$ of non-overlapping edge partitions, i.e. $E = \bigcup_{i=1}^{K}  \forall i,j : i \neq j \Rightarrow E_i \cap E_j = \emptyset$. 
 The $i^{th}$ partition is associated with a vertex set $V_i$, composed of the end points of its edges: $V_i = \{u : (u, v) \in E_i \vee (v, u) \in E_i\}$. 
 
 A \bladyg solution for edge partitioning in large dynamic graphs consists of two steps. 
 The first step computes the initial partitioning of the input graph. 
 Each vertex of each block maintains $\Block(u)$, which denote the block of the node $u$. 
 The second step consists in updating the partitioning according to the incremental changes. 
 Whenever a new edge $(u, v)$ is added to $G$, \bladyg first activates the \textit{M2W-mode} and assigns the edge $(u, v)$ to a selected block. 
 Block assignment is done by a $\masterCompute()$ operation that decides the block of each new edge considering predefined objective functions such as balance, communication efficiency and connectedness~\cite{DBLP:conf/europar/GuerrieriM15}. 
 The coordinator asks the worker that holds $u$ (respectively $v$) to compute the predefined objective functions inside $\Block(u)$ (respectively $\Block(v)$). 
 The result of this computation is sent to the coordinator that decides the block that will holds the edge $(u, v)$.
 Whenever an edge $(u, v)$ is removed from $G$, \bladyg asks all the workers to compute a \textit{repartitioning threshold} in order to decide if the partitioning of $G$ needs to be recomputed. 
 In each worker, the \textit{repartitioning threshold} is computed by a $\workerCompute()$ operation and sent to the coordinator. 
 The coordinator decides, by executing a $\masterCompute()$ operation, if a repartitioning of $G$ is needed or not. 

\subsection{Distributed maximal clique computation}

Given an undirected graph $G = (V, E)$, a clique is a subset of vertices $C \subseteq V$ such that every vertex in $C$ is connected to every other vertex in $C$ by an edge in $G$. 
A clique $C$ is called to be maximal if any proper superset of $C$ is not a clique. 
The problem of maximal clique enumeration (MCE) is to compute the set $ M(G)$ of maximal cliques in $G$. 
Considering the issue of dynamism, the problem of MCE in dynamic graphs~\cite{DBLP:conf/bigdata/XuCFB14} consists in incrementally update the set of maximal cliques for every graph update. 

\bladyg deals with the problem of MCE in dynamic graphs in the following way. 
Each edge of each block maintains $ID(v)$, $\adj(u)$, $M_u$ and $T_u$, which denote the identifier of $u$, the adjacent vertices of $u$, the set of maximal cliques of $u$ and a prefix-tree such that the root of $T_u$ is $u$ and each root-to-leaf path represents a maximal clique in $M_u$, respectively. 
We assume that adjacency list representation of the graph $G$, the set $V$ of vertices are ordered in ascending order of their IDs. 
We further define $\adj_{<}(u) = \{v : v \in \adj (u), ID(v) < ID(u)\}$ and $\adj_{>}(u) = \{v : v \in \adj (u), ID(v) > ID(u)\}$.
When an edge $(u, v)$ is inserted into $G$, \bladyg coordinator asks workers containing $u$ and $v$ to update the set of maximal cliques. 
Each of the workers of $u$ and $v$ executes a $\workerCompute()$ operation in order to remove existing maximal cliques that become non-maximal and insert maximal cliques that should be inserted. 
An existing maximal clique $C$ becomes non-maximal if $C$ contains either $u$ or $v$, and verifies $C \subset (\adj(u) \cap \adj(v)) \cup \{u, v\}$~\cite{DBLP:conf/bigdata/XuCFB14}. 
Maximal cliques that need to be added to the existing ones consists of new maximal cliques that contain ${u, v, w}$, for each $w \in ((\adj_{<}( u) \cap \adj_{<}( v)) \cup \{u\})$~\cite{DBLP:conf/bigdata/XuCFB14}. 
When an edge $(u, v)$ is deleted from $G$, \bladyg coordinator notifies workers containing the nodes $u$ and $v$ by the edge deletion. 
Workers that hold $u$ and $v$ execute a $\workerCompute()$operation that deletes all the existing maximal cliques that contain both $u$ and $v$, where such maximal cliques appear in $T_w$, where $w \in ((\adj_{<}( u) \cap \adj_{<}( v)) \cup \{u\})$~\cite{DBLP:conf/bigdata/XuCFB14}. 
Then, we generate all new maximal cliques that contain only $u$ or $v$, and insert them into $T_w$, where $w \in ((\adj_{>}( u) \cap \adj_{<}( v)) \cup \{v\})$ or $w \in ((\adj_{<}( u) \cap \adj_{<}( v)) \cup \{u\})$. 
A notification is sent to \bladyg coordinator when all the workers finish the update process. 

\section{Experiments}
\label{exp}
We have applied \bladyg framework to both the problem of distributed $k$-core decomposition in large dynamic graphs and the problem of partitioning of  dynamic graphs. 
We have performed a set of experiments to evaluate the effectiveness and efficiency of \bladyg framework on a number of different real and synthetic datasets. 
Implementation details of \bladyg can be found in the following link: \url{https://members.loria.fr/SAridhi/files/software/bladyg/}.
\subsection{Experimental environment}
We have implemented \bladyg on top of the \akka framework, a toolkit and runtime for building highly concurrent, distributed, resilient message-driven applications. 
In order to evaluate the performance of \bladyg, we used a cluster of $17$ \texttt{m3.medium} instances on Amazon EC2 ($1$ virtual $64$-bit CPU, $3.75$GB of main memory, $8$GB local instance storage).

\subsection{Experimental results}
\subsubsection{$k$-core decomposition of large dynamic graphs}

\begin{table*}
\centering
\caption{Experimental data}
\label{data1}
\scalebox{0.9}{
\begin{tabular}{|l|r|r|r|r|r|r|} \hline
\textbf{Dataset} & \textbf{Type} & \textbf{$\sharp$ Nodes} & \textbf{$\sharp$ Edges} &  $\oslash$ & \textbf{Avg. CC} & \textbf{Max(k)} \\ \hline
DS1 &Synthetic& 50,000 & 365,883&4 & 0.3929&42 \\ \hline
DS2 &Synthetic& 100,000 & 734,416&4 & 0.3908&46 \\ \hline
ego-Facebook  &Real& 4,039 & 88,234 & 8 &0.6055&115\\ \hline
roadNet-CA  & Real & 1,965,206 & 2,766,607 & 849& 0.0464& 3  \\ \hline
com-LiveJournal &Real&3,997,962 & 34,681,189 & 17&0.2843& 296  \\ \hline
\end{tabular}
}
\end{table*}

\begin{table*}
\centering
\caption{Experimental results}
\label{result1}
\scalebox{0.82}{
\begin{tabular}{|l|r|r|r|r|} \hline
\multirow{2}{*}{\centering \textbf{Dataset}}  		 &      \multicolumn{2}{c|}{\textbf{AIT (ms)}}		 &      \multicolumn{2}{c|}{\textbf{ADT (ms)}} \\  \cline{2-5}
        & \centering \textbf{inter-partition}& \centering \textbf{intra-partition} & \centering \textbf{inter-partition}&\textbf{intra-partition} \\ \hline
     DS1                  &42     &10      &32    &8      \\ \hline
     DS2                   &30     &10      &25    &8       \\ \hline
     ego-Facebook       &38     &15      &32    &10     \\ \hline
     roadNet-CA        &30     &12      &26    &10     \\ \hline
     com-LiveJournal    &256    &30      &205   &27       \\ \hline
\end{tabular}
}
\end{table*}


Since the goal is to compute $k$-core decomposition, the characteristic properties of our datasets (shown in Table~\ref{data1}) are the number of nodes, edges, the diameter, the average clustering coefficient and the maximum coreness.
We have used two groups of datasets: real-world ones, made available by the Stanford Large Network Dataset collection~\cite{snapnets}, and synthetic datasets, created by a graph generator based on the Nearest Neighbor model~\cite{Sala:2010:MGM:1772690.1772778}. 
Varying the input data enabled us to avoid biased results specific to a single dataset and thus to have a better interpretation of the results.

In order to simulate dynamism in each dataset, we consider two update scenarios.
For each scenario, we measure the performance of the system to update the core numbers of all the nodes in the considered graph after insertion/deletion of a constant number of edges:
\begin{itemize}
 \item In the \textit{inter-partition} scenario, we either delete or insert $1000$ random edges connecting two nodes belonging to \emph{different} partitions;
 \item In the \textit{intra-partition} scenario, we either delete or insert $1000$ random edges connecting two nodes belonging to \emph{the same} partition.
\end{itemize}

Table~\ref{result1} illustrates the  results obtained with both the real and the synthetic datasets.
For each dataset, we record the average insertion time (AIT) and the average deletion time (ADT) over the $1000$ insertions/deletions for both \textit{inter-partition} and \textit{intra-partition} scenarios.
To generate the results of Table~\ref{result1}, we randomly partition the graph dataset into 8 partitions. 
As shown in Table~\ref{result1}, we observe that in the \textit{intra-partition} scenario, the values of the average insertion/deletion time are much smaller than those in the \textit{inter-partition} scenario. 
This can be explained by the fact that the inserted/deleted edges in the \textit{intra-partition} scenario are internal ones. 
Consequently, the amount of data to be exchanged between the distributed machines in the case of internal edges is smaller, in most cases, than the amount of exchanged data in the case of edges of the \textit{inter-partition} scenario. 
During the $k$-core maintenance process after insertion/deletion of an internal edge, there is always the chance of not having to visit distributed workers/partitions other than the partition that holds the internal edge.

\begin{figure*}[t]
\centering
\includegraphics[width=0.9\textwidth]{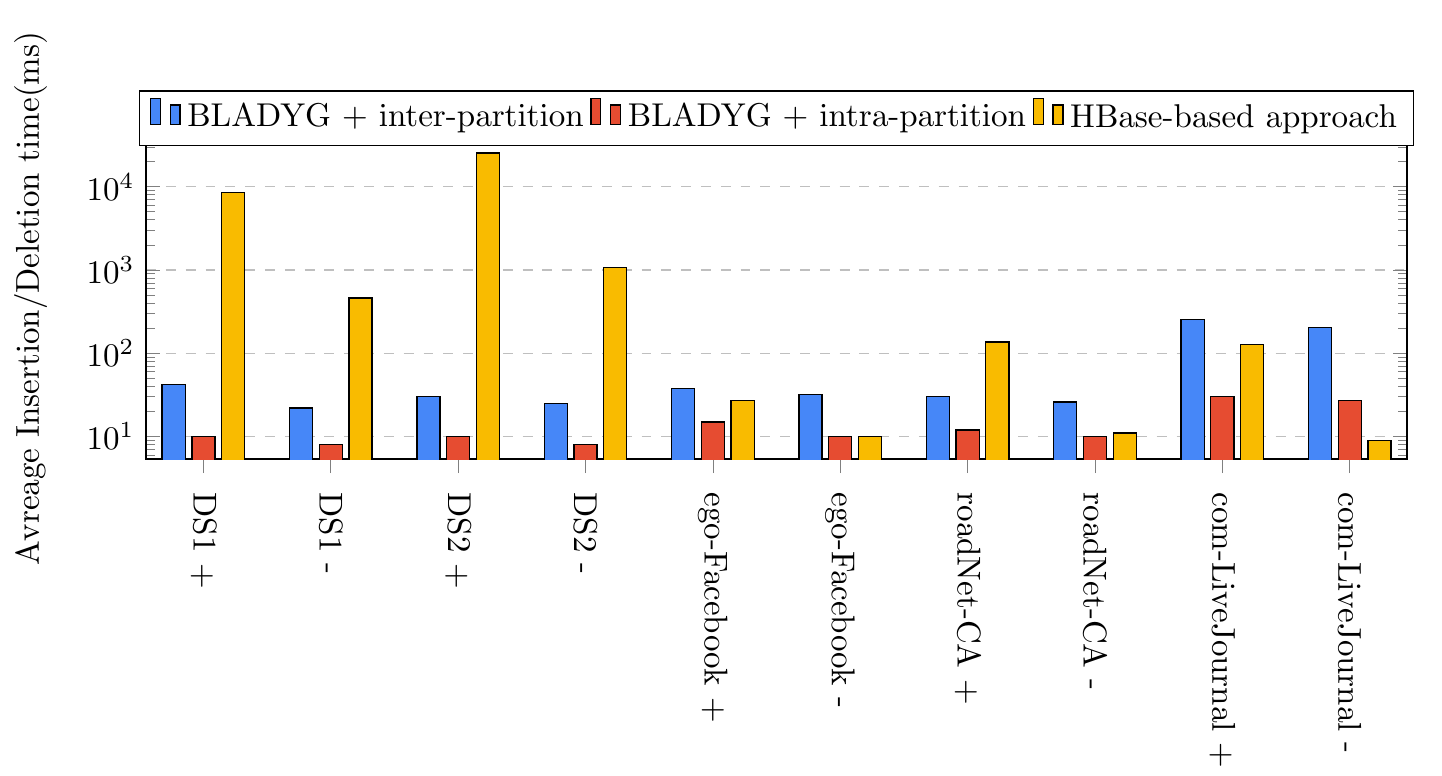}
\caption{Average insertion/deletion time}
\label{figinsertion}
\end{figure*}

Figure~\ref{figinsertion} presents a comparison of our \bladyg solution with the HBase-based approach proposed by Aksu et al. \cite{6702486} in terms of average insertion/deletion time. 
For our approach, we used $9$ \texttt{m3.medium} instances on Amazon EC2 (1 acting as a master and 8 acting as workers). 
For the HBase-based approach, we used $9$ \texttt{m3.medium} instances on Amazon EC2 (1 master node and 8 slave nodes). 
As stressed in Figure~\ref{figinsertion}, our approach allows much better results compared to the HBase-based approach for almost all datasets. 
It is noteworthy to mention that the presented runtime values of the HBase-based approach correspond to the maintenance time of only one fixed $k$ value core ($k = max(k)$ in our experimental study). 
This means that, for each dataset, the maintenance process of the HBase-based approach needs to be repeated $max(k)$ times in order to achieve the same results as our approach. 
\subsubsection{Partitioning of large dynamic graphs} 
The goal here is to evaluate the performance and the scalability of \bladyg solution for the partitioning of large and dynamic graphs. 
For our tests, we used the graph datasets described in Table~\ref{data1} and we considered three partitioning techniques (1) hash partitioning, (2) random partitioning and (3) \dynamicdfep, a previously published distributed partitioning algorithm \cite{ideas2016a}. \dynamicdfep is based on two main phases. The first phase consists of four steps: 
\begin{enumerate}
\item We randomly choose a single node for each of the desired partitions, and give it an initial amount of "funding" associated to that partition.
\item Each node will use its funding to the neighbors to try to "buy" additional edges. The partition will therefore buy the edges that are closer to the randomly chosen nodes and start getting bigger.
\item Since the initial amount of funding is insufficient for the partitions to cover the entire graph, additional funding is assigned to the partitions, in a manner inversely proportional to their size. A small partition (which may have been started far from the center of the graph) will receive more funding and therefore be more likely to grow than a larger partition.
\item Steps 2-3 are repeated until all edges have been bought by a partition.
\end{enumerate}
The second phase of \dynamicdfep deals with incremental changes by applying one of the supported update strategies. For our tests, we used the Unit-Based Update strategy (\ubupdate) described in \cite{ideas2016a}. 

In order to simulate dynamism in each dataset, we use only $90\%$ of the graph in the partitioning step and we insert the remaining $10\%$ in the update step. Each experiment is repeated five times and the numeric results in the following sections consists of the average over all runs. 

Tables~\ref{result2} and ~\ref{result3} illustrate the results obtained with both the real and the synthetic datasets.
For each dataset and for each partitioning method, we record the partitioning time (PT) and the update time (UT). The update time is computed for two different partitioning strategies: (1) \incrementalupdate and (2) \naivepart. The first update strategy consists in applying the used partitioning technique only on the incremental changes. The second update strategy is a naive partitioning technique that consists in destroying the old graph partitioning and all further information associated to the assignment and restarts from the scratch by running the used partitioning technique.

\begin{table*}
\centering
\caption{Experimental results using hash partitioning on BLADYG}
\label{result2}
\scalebox{1}{
\begin{tabular}{|l|c|c|c|} \hline
\multirow{2}{*}{\centering \textbf{Dataset}} & \multirow{2}{*}{\textbf{Partitioning time (s)}} &  \multicolumn{2}{c|}{\textbf{Update time (s)}} \\  \cline{3-4}
        &  &  \incrementalupdate&\naivepart \\ \hline
     DS1                  &18          &3    &19      \\ \hline
     DS2                   &43          &5    &40       \\ \hline
     ego-Facebook       &11        &2    &13     \\ \hline
     roadNet-CA        &180          &16    &193     \\ \hline
     com-LiveJournal    &209          &25   &227       \\ \hline
\end{tabular}
}
\end{table*}

\begin{table*}
\centering
\caption{Experimental results using random partitioning on BLADYG}
\label{result3}
\scalebox{1}{
\begin{tabular}{|l|c|c|c|} \hline
\multirow{2}{*}{\centering \textbf{Dataset}}  &   \multirow{2}{*}{\textbf{Partitioning time (s)}} &  \multicolumn{2}{c|}{\textbf{Update time (s)}} \\  \cline{3-4}
        &  &  \incrementalupdate&\naivepart \\ \hline
     DS1                  &21          &3    &20      \\ \hline
     DS2                   &36          &6    &42       \\ \hline
     ego-Facebook       &12        &2    &10     \\ \hline
     roadNet-CA        &171          &21    &202     \\ \hline
     com-LiveJournal    &211          &27   &232       \\ \hline
\end{tabular}
}
\end{table*}

\begin{table*}
\centering
\caption{Experimental results using \dynamicdfep on BLADYG}
\label{result4}
\scalebox{1}{
\begin{tabular}{|l|c|c|c|} \hline
\multirow{2}{*}{\centering \textbf{Dataset}}  &   \multirow{2}{*}{\textbf{Partitioning time (s)}} &  \multicolumn{2}{c|}{\textbf{Update time (s)}} \\  \cline{3-4}
        &  &  \ubupdate&\naivepart \\ \hline
     DS1                  &30          &3    &32      \\ \hline
     DS2                   &56          &5   &62       \\ \hline
     ego-Facebook       &80        &2    &91     \\ \hline
     roadNet-CA        &254          &31    &321     \\ \hline
     com-LiveJournal    &509          &34  &572       \\ \hline
\end{tabular}
}
\end{table*}

As shown in Tables~\ref{result2},~\ref{result3} and~\ref{result4}, we observe that, for all partitioning methods, \bladyg results using \incrementalupdate strategy are much better than those using \naivepart strategy. 
This can be explained by the fact that \bladyg allows to process only incremental changes without restarting the partitioning from the scratch.


\section{Conclusions}
\label{conc}
This paper deal with the problem of graph processing in large dynamic networks. 
We presented \bladyg framework, a block-centric framework that addresses the issue of dynamism in large scale graphs. 
The presented framework can be used not only to compute common properties of large graphs but also to maintain the computed properties when new edges and nodes are added or removed.
We implemented \bladyg on top of \akka, a framework for building highly concurrent, distributed, and resilient message-driven applications. 
We applied \bladyg to two different problems: (1) distributed $k$-core decomposition in large dynamic graphs and (2) partitioning of large dynamic graphs. 
By running some experiments on a variety of both real and synthetic datasets, we have shown that the performance and scalability of the proposed framework are satisfying for large-scale graphs. 

In the future work, we aim at studying data communications, networking and scalability of \bladyg framework with respect to the number of distributed machines. 

{\normalsize

\bibliographystyle{abbrv}
\bibliography{biblio}  
}
\end{document}